\documentclass[twocolumn,twocolappendix, times]{aastex701}
\usepackage{amsmath} 
\usepackage{amssymb} 
\usepackage{graphicx, graphics}
\usepackage{xcolor}
\usepackage{ragged2e}
\usepackage{soul}

\usepackage[flushleft]{threeparttable}
\usepackage[T1]{fontenc}
\usepackage{enumitem}
\usepackage{newtxtext,newtxmath,ulem}
\usepackage{multirow}

\usepackage{hyperref}
\hypersetup{colorlinks=true, 
    linkcolor=blue, 
    filecolor=magenta, 
    urlcolor=blue, 
    citecolor=blue, 
    pdfauthor={Kartick Sarkar}, 
    pdftitle={Fermi Bubbles Review} 
}

\newcommand{\msun}{M$_{\odot}$ }

\newcommand{\pcc}{cm$^{-3}$ }

\newcommand{\psec}{s$^{-1}$ }
\newcommand{\ergps}{erg \psec}
\newcommand{\mpy}{\msun yr$^{-1}$ }

\renewcommand{\deg}{^\circ}

\setstcolor{black}

\begin{document}

\title{Can the $\sim 5$ kpc starburst ring in the Milky Way create the eROSITA bubbles?}

\author[0000-0002-1497-4645]{Mukesh Singh Bisht}
\affiliation{Institute of Astronomy, National Tsing Hua University, Hsinchu 30013, Taiwan}
\affiliation{Raman Research Institute, Bengaluru, India}
\email[show]{mukesh@gapp.nthu.edu.tw}  

\author[0000-0002-7767-8472]{Kartick C. Sarkar} 
\affiliation{Raman Research Institute, Bengaluru, India}
\email[show]{kcsarkar@rrimail.rri.res.in}

\begin{abstract}
eROSITA bubbles (eRBs) in our Galaxy are thought to be due to the energy injection by the central star-forming region or the central supermassive black hole. Identifying the origin of these bubbles is crucial in constraining the origin of the Fermi Bubbles and the nature of the Galactic feedback. However, some recent opinions suggest that the eRBs may not even have been produced by any central source, but rather a result of a star-formation ring at $R\sim 5$ kpc. In this letter, we perform detailed hydrodynamic simulations to refute that claim and show that such a ring of star formation alone is not able to produce the symmetric eROSITA bubbles - one needs a compact energy injection from the Galactic Center to produce these bubbles. 
Our study, therefore, suggests that the eROSITA bubbles truly represent the nuclear outflow from our Galaxy.

\end{abstract}



\section{Introduction}
\label{sec:introduction}

All-sky observations of the Milky Way in multi-wavebands have revealed the presence of large-scale bipolar structures extending above and below the Galactic plane. In gamma rays, the most prominent features are the Fermi Bubbles (FBs) \citep{Su2010, Ackerman2014}. 
These structures appear as two lobes symmetrically located about the Galactic center and extend to Galactic latitudes of roughly $|b| \lesssim 50^{\circ}$ and galactic longitudes of $|l| \lesssim 20^{\circ}$. Their sharp edges and relatively uniform surface brightness suggest that they were produced by a powerful, energetic event at the Galactic Center (GC). In X-ray, eROSITA revealed large-scale X-ray structures encompassing the gamma-ray bubbles in both the Galactic hemispheres, now referred to as the eROSITA Bubbles \citep{Predehl2020}. The north-eastern part of the eROSITA Bubbles (eRBs) is also observed by the ROSAT all-sky survey and is known as the North Polar Spur (NPS)/Loop-I \citep{Snowden1997}. 
These X-ray features likely trace shock-heated gas surrounding the gamma-ray emitting regions. The presence of coherent bipolar structures in both gamma-ray and X-ray maps strongly suggests a large-scale outflow originating from the inner regions of the Milky Way.

The physical origin of these structures (FBs/eRBs) remains uncertain. Proposed explanations broadly fall into two categories. One possibility is that the bubbles were generated by past activity of the central supermassive black hole, Sagittarius A$^*$ \citep{Yang2012, Zubovas2012, Yang2018, Mondal2022, Sarkar2023}, analogous to outbursts observed in active galactic nuclei. 
A second possibility is that the bubbles are the result of large-scale winds driven by star formation in the central region of the Galaxy, where the combined effects of stellar winds and supernova explosions generate a powerful outflow \citep{Crocker2015, Sarkar2015, Sarkar2017}. While several authors connect the FBs to the eRBs in a unified model \citep{Sarkar2015, Sarkar2019, Yang2022}, others do not. The uncertainty in connecting the eRBs to the FBs has been the uncertain distance of the eRBs. While the superposition of eRBs and FBs and extinction-based distance indicators suggests a common origin \citep{Lallement2016, Lallement2022}, the starlight polarization-based distance indicators suggest a separate origin for the FBs and eRBs (\citealt{Spoelstra1972, Vidal2015}, also see \citealt{Sarkar2024} for a detailed review). More recently, based on radio polarization in the eRBs, \citealt{ZhangH2024} and \citealt{Churazov2024} suggested that the eRBs may be unrelated to the central energy injection and could simply be overlapping hot gas plumes rising from the $\sim 5$ kpc star-forming ring in our Galaxy \citep{Elia2022}. Given that the hot plumes are under-dense, they further proposed that the metallicity inside these plumes of gas must be much higher than solar for them to be bright in $\sim$ keV X-ray bands. If true, this origin scenario of the eRBs would affect the unified origin of the FBs and eRBs and alter the understanding of the nuclear feedback from our Galaxy.

In this work, we investigate whether stellar feedback associated with the $\sim 5$ kpc star-forming ring can give rise to the large-scale structures observed in the X-ray all-sky maps, or whether energy injection confined to the central region of the Galaxy is required to reproduce their morphology. Using detailed hydrodynamical simulations, we compare the morphology and X-ray intensity maps of both these scenarios and find that a strong central energy injection is an essential component of the Galactic feedback, without which the eRBs would not be present in our Galaxy. In the next sections, we describe our methodology and results in detail.

\section{Simulation set up}
\label{sec:simulation_setup}
We perform our hydrodynamical simulations in three-dimensional Cartesian coordinates using the publicly available code \textsc{pluto} \citep{Mignone2007}. Our initial Galactic density and potential setup is similar to that of \citealt{Sarkar2015a}, consisting of a rotating interstellar medium (ISM) and a non-rotating, isothermal circumgalactic medium (CGM), embedded in the gravitational potential of a cored Navarro–Frenk–White halo \citep{Navarro1997} and a Miyamoto–Nagai disk \citep{Miyamoto1975}. The Galactic parameters are provided in Table \ref{tab:galactic-parameters}. The ISM is initialized at a temperature of $T_{\rm ism} = 4 \times 10^{4}$ K (thermal plus dynamic temperature), while the CGM is set to $T_{\rm cgm} = 3 \times 10^{6}$ K. Our initial density distribution is in a steady-state equilibrium with the background potential (see \citealt{Sarkar2015a} for further details).

We define the simulation domain to extend from $-12$ kpc to $12$ kpc in both the $x$ and $y$ directions, and from $0$ to $15$ kpc in the $z$ direction. We adopt a uniform grid in the $x$ and $y$ directions with $512$ cells each, corresponding to a spatial resolution of $\approx 47$ pc. In the $z$ direction, the grid is uniform with $100$ grid points (spatial resolution of $2$ pc) from $0$ to $200$ pc and then with $412$ grid points up to $15$ kpc (spatial resolution of $\approx 36$ pc). Such grids allow us to properly resolve the ISM region (scale height of $\sim 200$ pc and a scale radius of $\sim 4$ kpc) in all directions. 
We choose a reflective boundary condition at the inner ($z=0$) boundary in the $z$ direction. At all other boundaries, the pressure, density, and velocity are fixed to their initial equilibrium values to maintain no loss of energy and mass through the boundaries.
We adopt the \textit{power-law} cooling prescription implemented in \textsc{pluto}. To construct the cooling function, we computed radiative cooling rates with $1Z_\odot$ and $0.3Z_\odot$ metallicities using the radiative transfer code \textsc{cloudy-v17} \citep{Ferland2017}. The resulting cooling curves were fitted with broken power laws over nine temperature bins spanning $10^4$--$10^9$ K. In each temperature bin, we express the cooling function as $\Lambda(T) = \Lambda_0\times T^\alpha$.
The normalization ($\Lambda_0$) in the first temperature bin is $2.2\times10^{-34}$ \ergps cm$^3$ K$^{-\alpha}$ for $1Z_\odot$ and $8.5\times10^{-37}$ \ergps cm$^3$ K$^{-\alpha}$ for $0.3Z_\odot$ metallicity. The power-law index ($\alpha$) values are provided in Table \ref{tab:cooling-parameters}. Cooling rates for intermediate metallicities are obtained through one-dimensional interpolation between these two cooling curves at a given temperature. 
We do not allow the ISM ($R = \sqrt{x^2+y^2} < 12$ kpc and $z < 2$ kpc) to cool unless it is shock-heated \footnote{Ideally, the ISM would be heated and stirred by individual supernovae to keep it at a thermal temperature of $\approx 10^4$ K, but a dynamical temperature of $4\times 10^4$ K (considering turbulence). Our assumed temperature in the initial setup includes the dynamical pressure due to turbulence. Therefore, we use the radiative cooling restriction in the ISM to avoid artificial cooling of the dynamical temperature.}.
The metallicities of the injected material and the CGM are set to $1\, Z_{\odot}$ and $0.3Z_{\odot}$, respectively. 

To test whether the eROSITA bubbles are produced due to the star-forming ring at $\sim 5$ kpc or a central source, we conduct two sets of runs. In the first run, we launch outflows solely due to disk-wide star formation, with a peak star formation around $\sim 5$ kpc. We refer to this run as `DI' (distributed injection). In the second run, along with disk-wide stellar energy injection, we also inject energy at the central region of the Galaxy. We refer to this run as `DI+CI' (distributed injection + central injection).

\begin{table*}[]
    \centering
    \caption{Galactic parameters considered for our simulations.}
    \begin{tabular}{|c|c|c|c|c|c|c|c|}
    \hline
        parameters & $M_{\rm vir}$ & $M_{\rm disc}$  &  $T_{\rm cgm}$  & $r_{\rm vir}$ &  $Z_{\rm cgm}$ & $Z_{\rm ism}$  \\
       \hline
      values & $10^{12}$ \msun & $5\times 10^{10}$ \msun & $3 \times 10^6$ K      & 258 kpc  & 0.3 $Z_\odot$ & 1 $Z_\odot$ \\
       \hline 
    \end{tabular}
    \label{tab:galactic-parameters}
\end{table*}

\begin{table*}
\centering
\caption{Power-law cooling indices, $\alpha$ ($\Lambda(T) = \Lambda_0 T^{\alpha}$) in different temperature ranges (in log$_{10}$) for $1Z_\odot$ and $0.3Z_\odot$ metallicity. The $\Lambda_0$ is set for the first temperature bin and is equal to $2.2\times10^{-34}$ \ergps cm$^3$ K$^{-\alpha}$ for the $1Z_\odot$ cells and $8.5\times10^{-37}$ \ergps cm$^3$ K$^{-\alpha}$ for the $0.3Z_\odot$ cells.}
\begin{tabular}{|l|c|c|c|c|c|c|c|c|c|}
\hline
log(T [K]) &
$4.0-4.4$ &
$4.4-4.95$ &
$4.95-5.4$ &
$5.4-5.6$ &
$5.6-5.88$ &
$5.88-6.18$ &
$6.18-6.6$ &
$6.6-7.55$ &
$7.55-9.0$ \\
\hline
$1\,Z_\odot$ &
$2.60$ &
$1.55$ &
$0.10$ &
$-2.00$ &
$-0.40$ &
$-0.15$ &
$-1.60$ &
$-0.15$ &
$0.45$ \\

$0.3\,Z_\odot$ &
$3.10$ &
$1.20$ &
$-0.10$ &
$-1.90$ &
$-0.40$ &
$-0.15$ &
$-1.35$ &
$0.18$ &
$0.48$ \\
\hline
\end{tabular}
\label{tab:cooling-parameters}
\end{table*}

\begin{figure}
    \centering
    \includegraphics[width=0.45\textwidth]{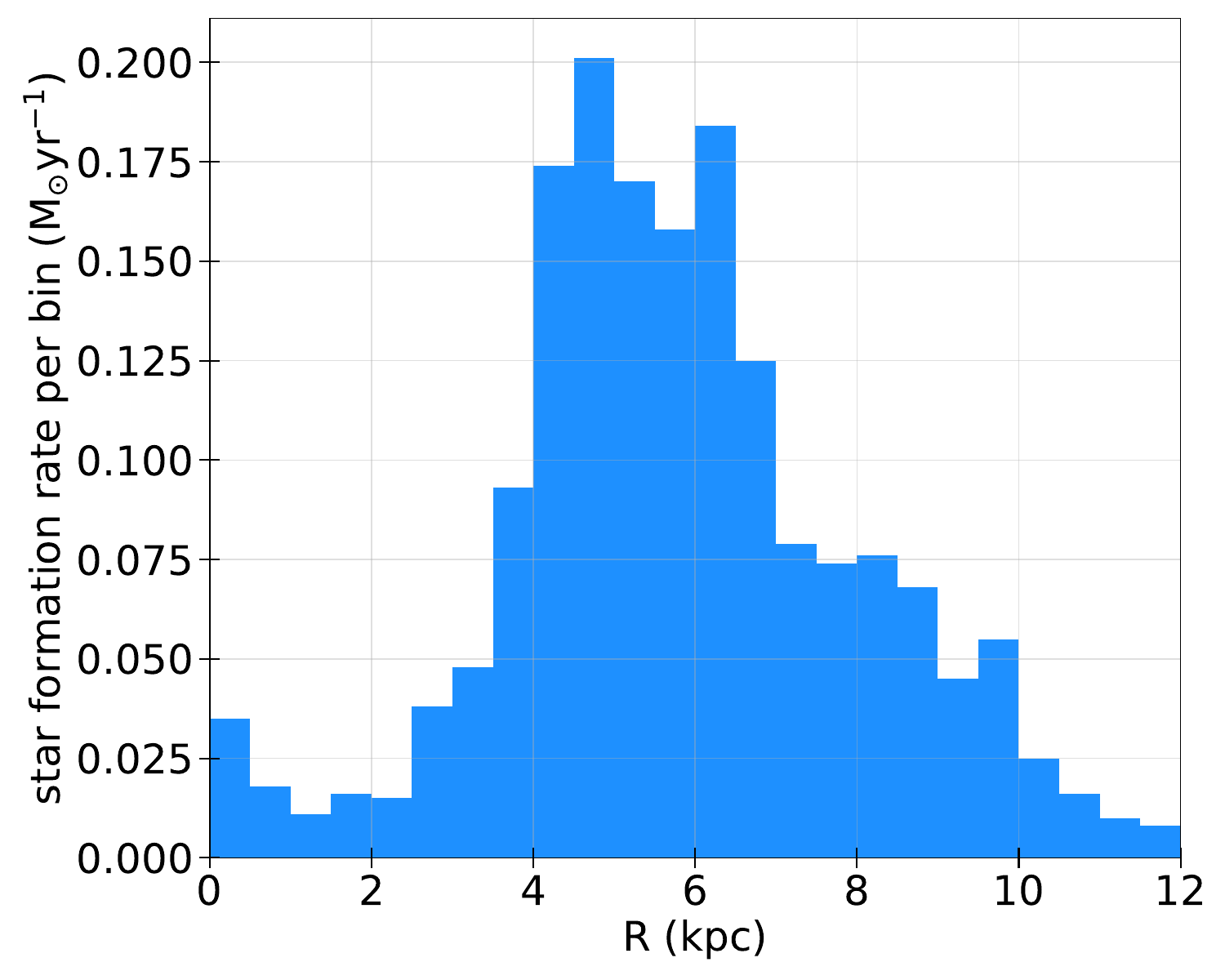}
    \caption{Observed star formation rate in each bin with a width of $0.5$ kpc as a function of $R$ (from \citealt{Elia2022}). The peak star formation is around $\sim 4-5$ kpc (star-forming ring), which is thought to be responsible for the eROSITA bubbles.}
    \label{fig:sfr_R}
\end{figure}

\begin{figure*}
    \centering
    \includegraphics[width=\textwidth]{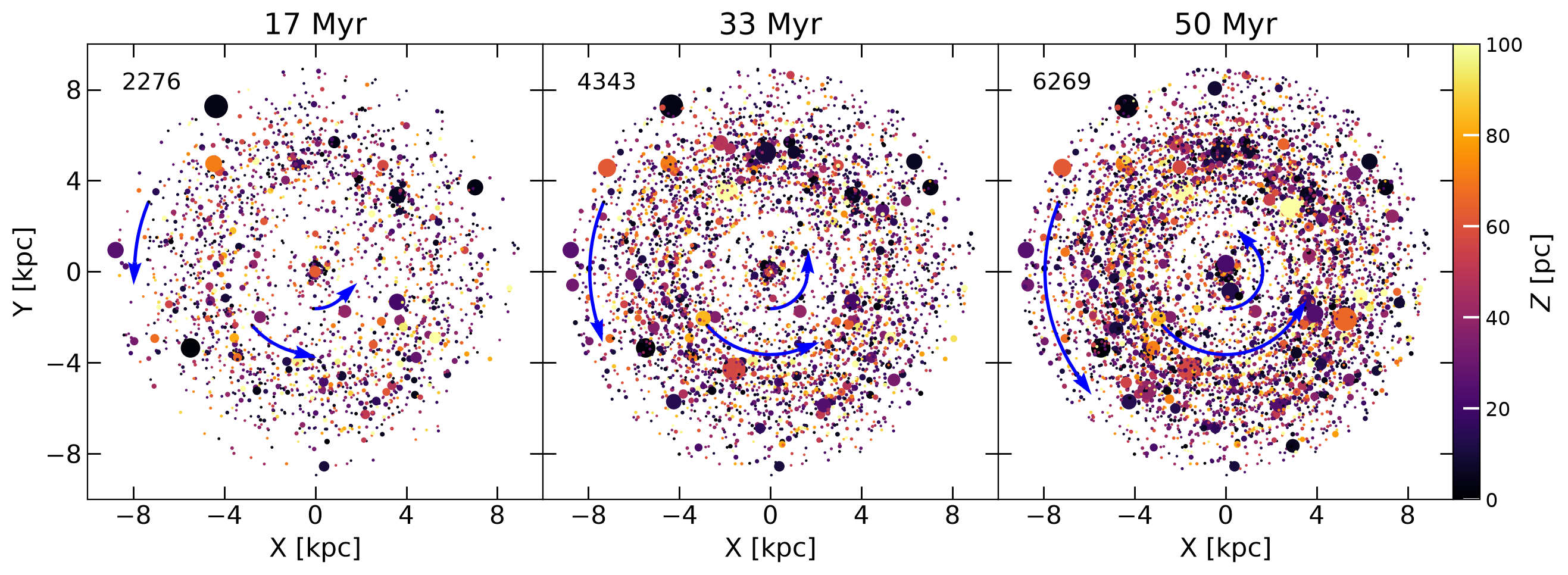}
    \caption{Birth positions $(x,y)$ of the star clusters formed until 17, 33, and 50 Myr are marked, with larger (smaller) points showing massive (low-mass) star clusters. Markers are color-coded according to the birth $z$ coordinates of the respective star clusters. Trajectories of 3 star clusters are shown in blue. The arrows show the direction of motion of star clusters. The total number of star clusters (indicated by the numbers in each panel) increases with time, as shown in each sub-panel. A `ring' of star clusters is formed around $R\sim 4-5$ kpc, where the star formation peaks (see Fig. \ref{fig:sfr_R}).}
    \label{fig:SC_orbits}
\end{figure*}

\begin{figure}
    \centering
    \includegraphics[width=0.9\linewidth]{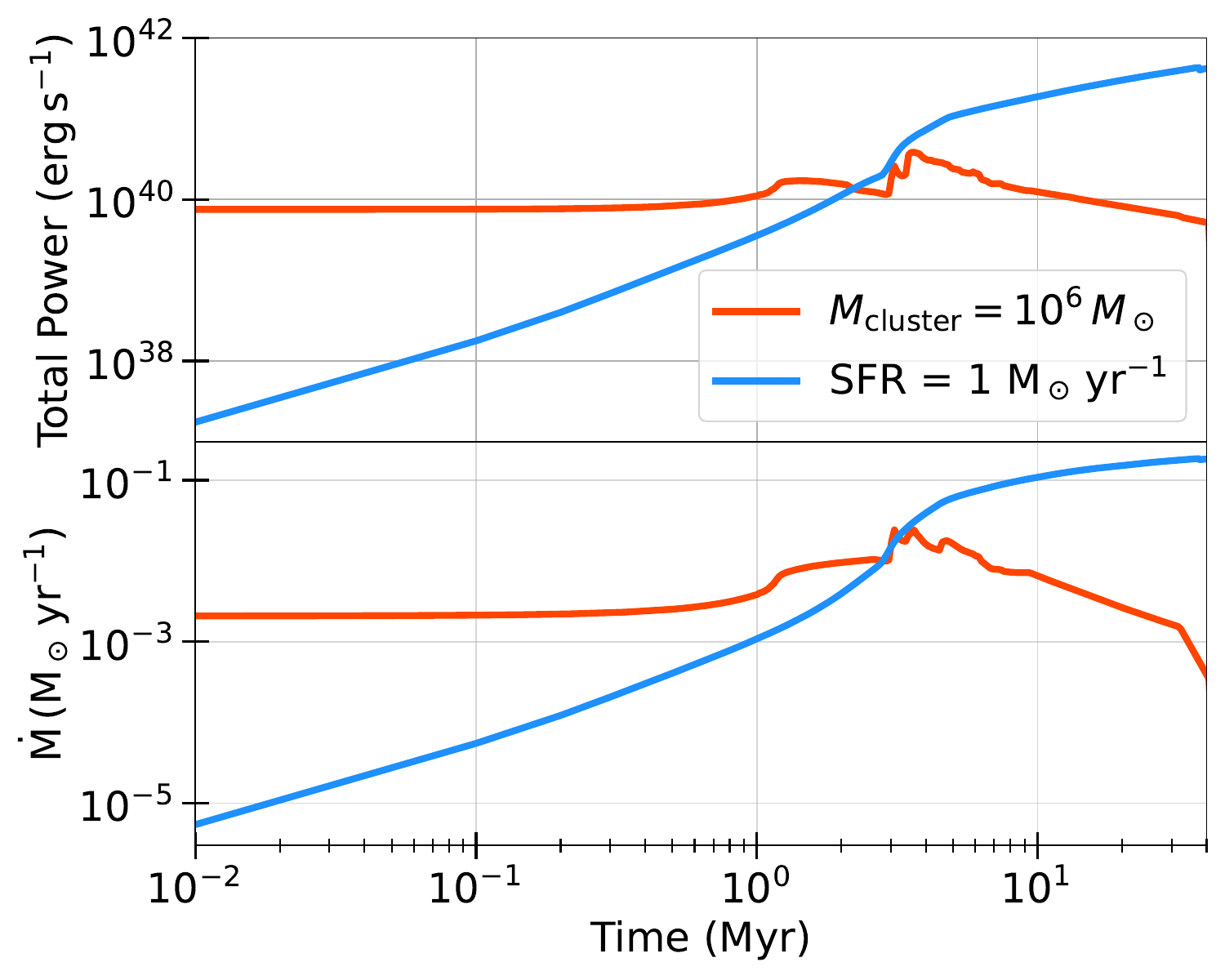}
    \caption{Energy and mass output rates of a young star cluster of mass $10^6$ M$_\odot$ (red lines) and a star-forming region with a star formation rate of $1$ \mpy (blue) as a function of time. The rates are generated using the \textsc{starburst99} code.}
    \label{fig:mass_en_inj}
\end{figure}

\begin{figure*}
    \centering
    \includegraphics[width=0.9\textwidth]{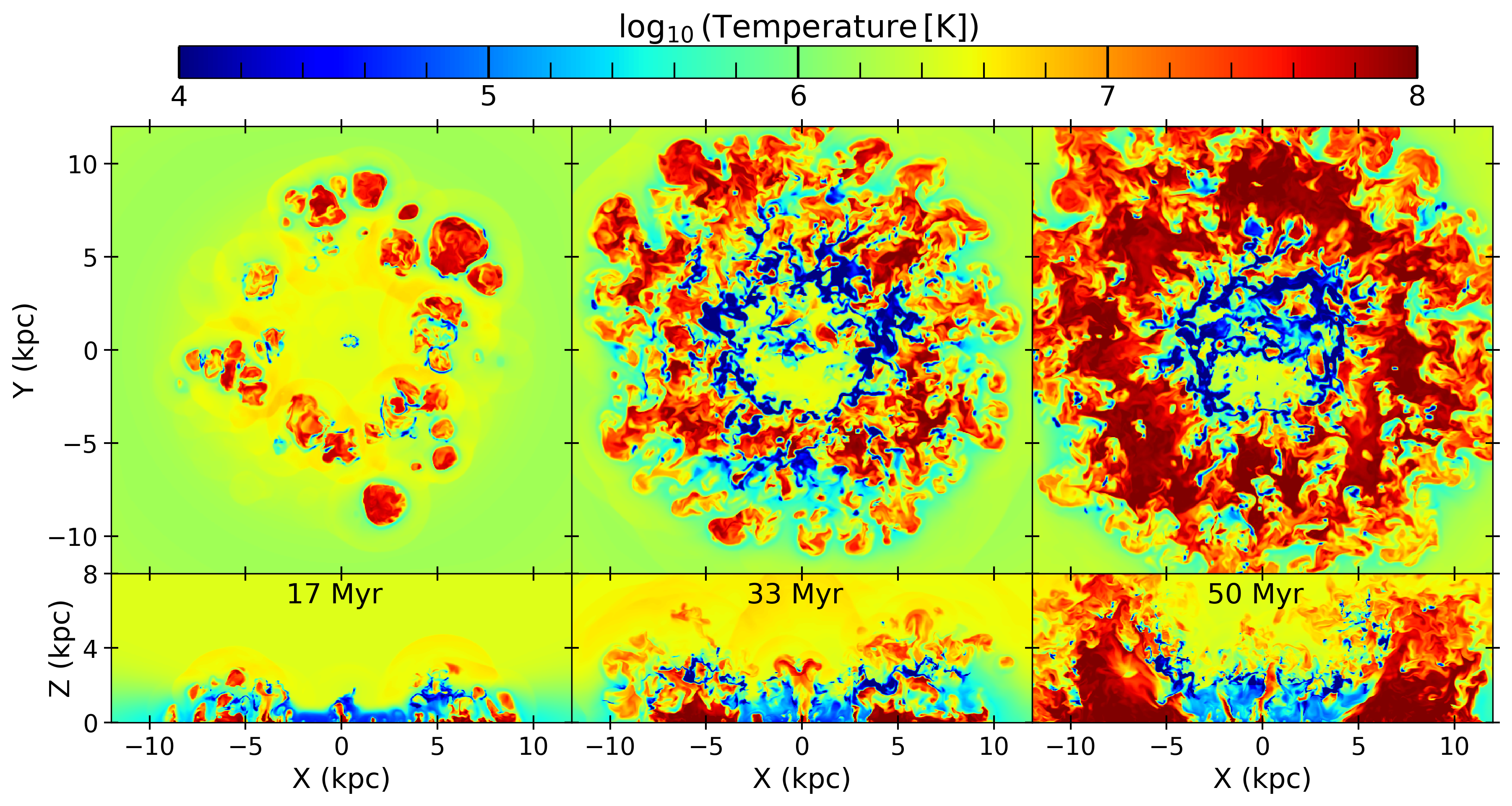}
    \caption{Temperature maps of the 'DI' simulation (without the central energy injection). \textit{Top Panel:} face on view at $17$, $33$, and $50$ Myr at $z = 2$ kpc. With time (left to right), a larger number of star clusters are generated, resulting in hot (red) bubbles. A ring-like structure is seen around $\approx 5$ kpc, which corresponds to the peak in the SFR as shown in Fig. \ref{fig:sfr_R}. \textit{Bottom Panel:} edge-on view at $17$, $33$, and $50$ Myr at $Y=0$ kpc. Two horn-like chimneys rising from $X\sim 5$ kpc correspond to the overlapping hot bubbles from individual star clusters.}
    \label{fig:time_evol_temp_DI}
\end{figure*}

\begin{figure*}
    \centering
    \includegraphics[width=0.9\textwidth]{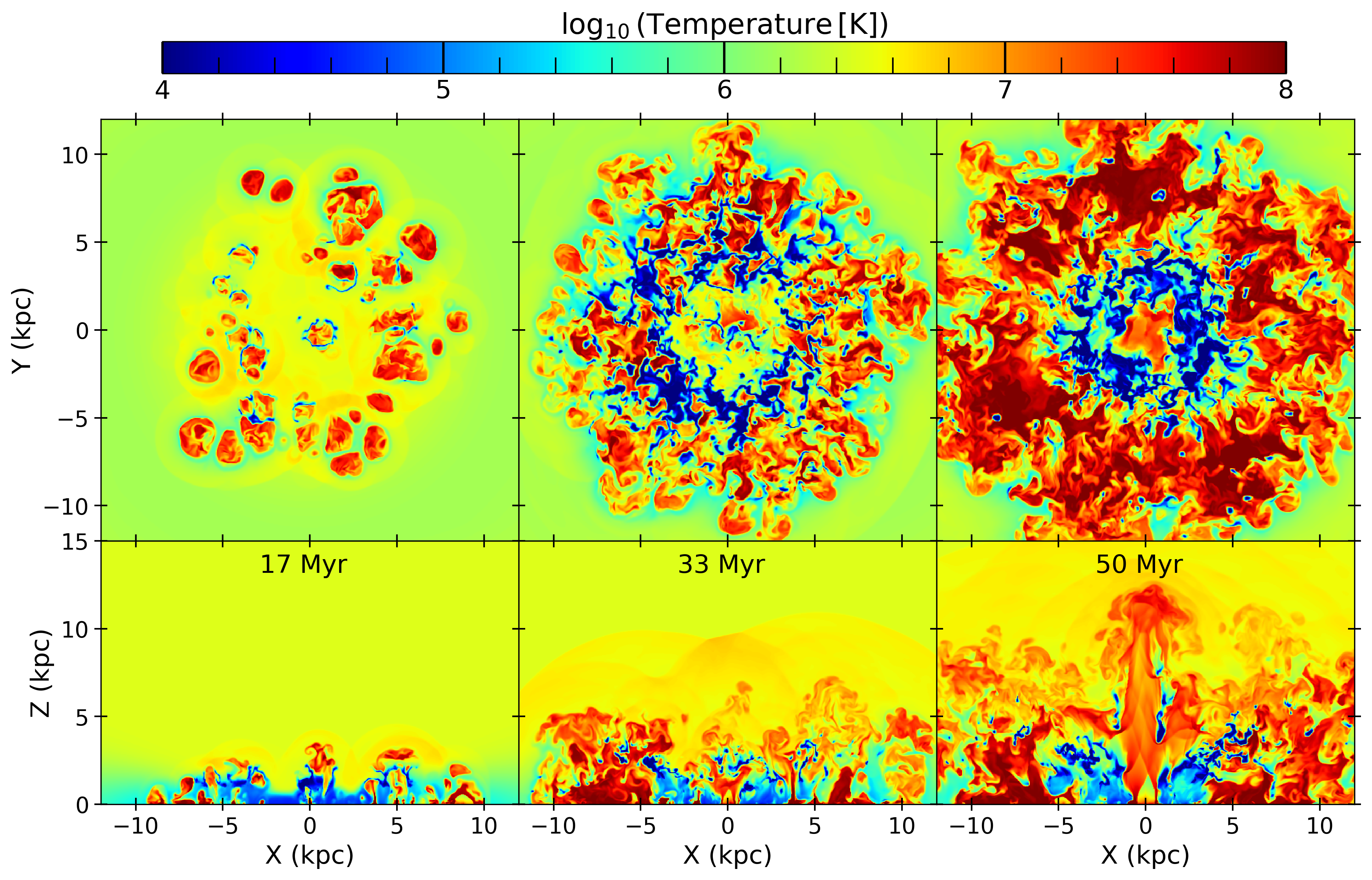}
    \caption{Same as Figure \ref{fig:time_evol_temp_DI}, but for the distributed + central injection (DI+CI) case.}
    \label{fig:time_evol_temp_CI}
\end{figure*}

For stellar feedback, the energy and mass injection rate depend on the star formation rate (SFR). Observations show that star-forming regions in the Milky Way are not concentrated at the Galactic center but are instead distributed throughout the Galactic plane, with a peak in SFR at $\sim 4\hbox{-}5$ kpc from the Galactic center (e.g., \citealt{Elia2022}, also see Fig. \ref{fig:sfr_R}). 
We assume that all star formation in each bin results in the formation of star clusters. We sample the mass of star clusters from a power-law cluster mass function with index $-2$, with lower and upper mass limits of $10^3$ and $10^6$ M$_\odot$, respectively \citep{Zhang1999,Chandar2026}. For a given SFR and cluster mass $\rm M_{\rm cl}$, we define a time interval $\tau=M_{\rm cl}/\rm SFR$ during which only one cluster forms. For example, an SFR of $0.1$ M$_\odot$ yr$^{-1}$ and a cluster mass of $10^5$ M$_\odot$ corresponds to $\tau=1$ Myr. A star cluster is randomly generated within this time interval. This procedure is applied to all SFR bins with their respective SFRs and cluster masses for the whole simulation period. After the formation of a cluster in a given radial bin, the mass of the next cluster in the same bin is resampled from the cluster mass function, resulting in a distribution of cluster masses within each radial bin over time.
Once a star cluster of a given mass is formed, it is randomly positioned within the corresponding radial bin, and its position subsequently evolves according to the Galactic rotation ($v_\phi(R)$, calculated from the assumed gravitational potential) at that radius. The $x$ and $y$ coordinates of the cluster are randomly generated within the bin, while the $z$ coordinate is drawn from a Gaussian distribution with zero mean and a standard deviation of $50$ pc.
This is to ensure that the cluster formation qualitatively follows the ISM density distribution in the $z$-direction. 
In Fig. \ref{fig:SC_orbits}, each sub-plot shows the birth positions $(x,y)$ of the star clusters formed until 17, 33, and 50 Myr, respectively. The colors of the scatter points indicate the birth $z$ coordinates of the star clusters. The total number of star clusters (2276, 4343, 6269 until 17, 33, and 50 Myr, respectively) shown within each sub-plot increases with time. The trajectories of 3 star clusters are shown in blue for illustration, with arrows showing their direction of motion. The size of points varies according to the mass of star clusters formed, with the biggest size corresponding to $\sim 10^6$ \msun. 
Mass and energy are injected based on the cluster mass and its age from its formation time in the simulation. The mass and energy injection rates are obtained from the \textsc{starburst99} code \citep{Leitherer1999} for a given star cluster mass ($10^6$ \msun) and for a given SFR ($1$ \mpy) and then scaled for the clusters in our simulations (see Fig. \ref{fig:mass_en_inj}). 
The mass and energy are injected into a spherical region of radius $50$ pc centered on the instantaneous location of the star cluster. 

In simulation DI, we inject mass and energy solely from star clusters, as described above. In simulation DI+CI, in addition to star cluster feedback, we also inject mass and energy at the Galactic center to model the eROSITA bubbles. As shown by \citealt{Sarkar2015}, the timescale for the emergence of these bubbles is $\approx 27$ Myr; accordingly, we start injecting mass and energy at the Galactic center at $23$ Myr to recreate the eRBs at $50$ Myr. Each simulation is evolved for a total duration of $50$ Myr, by which time the stellar cluster outflows are expected to reach a statistical equilibrium since the dynamical time of the stellar outflows in the inner CGM is $t_{\rm dyn} \lesssim 10$ kpc/$c_s(T_{\rm cgm}) \sim 40$ Myr. Practically, the stellar outflows may produce a shock with $Mach \gtrsim 1$, and therefore, the dynamical time of these outflows would be even smaller. The mass and energy injection rates for the central injection (CI) are shown by the blue curves in Figure \ref{fig:mass_en_inj}, linearly scaled to $0.5$ \mpy. For central injection (CI), we inject mass and energy in a spherical region of radius $50$ pc at the GC. As has been shown in \citealt{Sarkar2015}, the exact value of this injection radius at the GC does not affect the dynamics or shape of the overall outflow.

\begin{figure*}
    \centering
    \includegraphics[width=\textwidth]{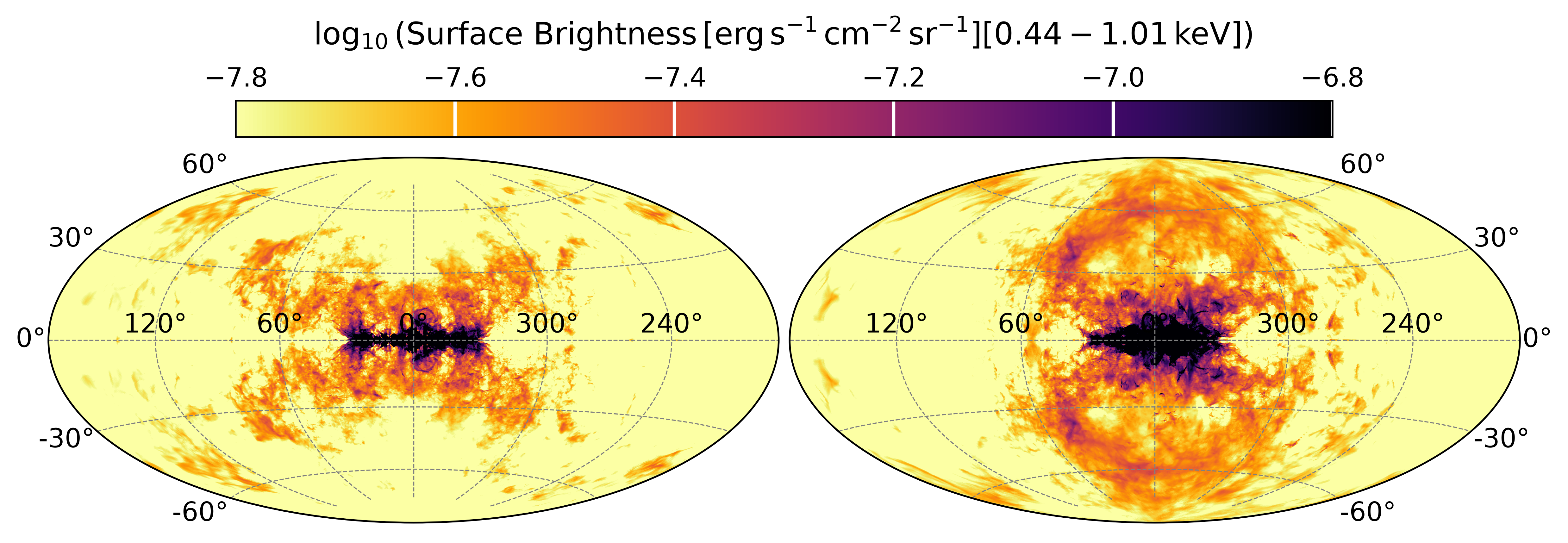}
    \caption{X-ray surface brightness maps of two scenarios of star formation in the MW. One, without the presence of the central energy injection (left panel), and second, with the presence of a central energy injection corresponding to an SFR $= 0.5$ \mpy (right panel). The absence of an eROSITA bubble in the first case clearly showcases that the $\sim 5$ kpc star formation ring in the MW is not able to produce the eRBs, and a central injection is necessary.}
    \label{fig:bubble}
\end{figure*}

\begin{figure*}
    \centering
    \includegraphics[width=\textwidth]{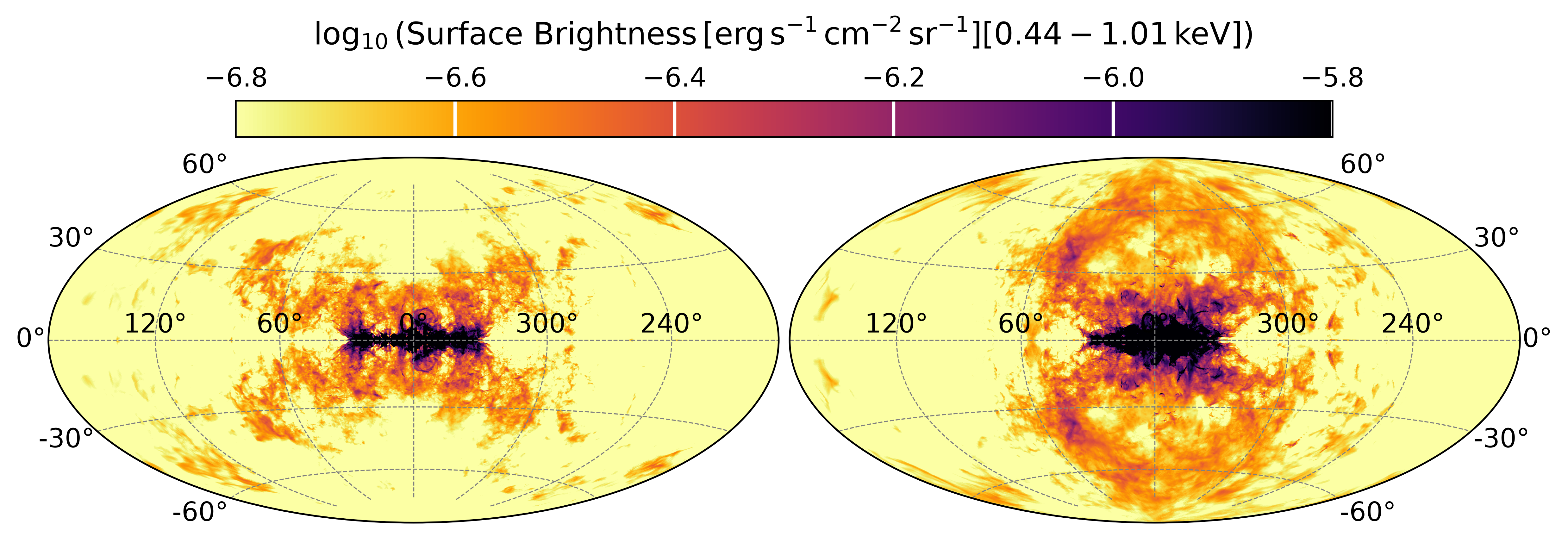}
    \caption{Effect of metallicity on the projected X-ray sky map: The metallicity of the injected gas (traced by a tracer value in \textsc{pluto}) is artificially boosted by 10 times while making the projection maps. This is to ensure a high metallicity injection by the supernovae. We do not find any evidence of the eROSITA bubbles in the DI case (left panel) even after such a metallicity enhancement, while eROSITA bubbles are still present in the DI+CI case (right panel). This implies that the eROSITA bubbles are not simple low-density chimneys with high metallicity.}
    \label{fig:metal_var}
\end{figure*}

\section{Results \& Conclusion}
\label{sec:result}

Figure \ref{fig:time_evol_temp_DI} shows the time evolution of temperature in our `DI' run \footnote{Full movies of the simulations are available at \url{https://youtu.be/wdTZBoKuExo}}. 
As we see, with time, more star clusters form, producing distributed outflows that appear as hot and red regions. 
The strong outflows are generated at $\sim \pm 5$ kpc around the star formation ring, due to the high cluster formation rate at these locations and the overlapping hot bubbles from individual clusters. These hot regions, when intersected along the line of sight, will contribute to the total X-ray emission, along with that from the extended CGM. This is also evident in the density map shown in Fig. \ref{fig:time_evol_density_DI} and \ref{fig:time_evol_density_CI}.
We see that chimneys of hot gas ($T\sim 10^7-10^8$ K) extend all the way to $\approx 10$ kpc in the radial direction and up to $8$ kpc in the vertical direction and have a width of about $\sim 5$ kpc, as a result of non-vanishing star formation at those radii (Fig. \ref{fig:sfr_R}). Although the rise of the horn-like hot gas chimney structures at the $\sim 4-5$ kpc region is correctly predicted by \citealt{Churazov2024}, we do not find any sharp edges to these horns. Furthermore, the density of these hot chimneys is $\sim 10^{-5}-10^{-4}$ \pcc, making these chimneys unsuitable for strong X-ray emission at $\sim 1$ keV energies. Therefore, these chimneys are unlikely to represent the sharply edged eRBs. Shocks around these chimneys, however, have a temperature ($\sim 5\times 10^6$ K)  and density ($\sim 10^{-3}$ \pcc) that are suitable for strong X-ray emission at $\sim 1$ keV. To understand the overall X-ray emission intensity and morphology, one has to consider the full projection effects. 

For both runs, we generated the projected surface brightness (SB) map in ROSAT band 4, $0.44\hbox{--}1.01$ keV, using the projection code \textsc{pass} \citep{Sarkar2017}. \textsc{pass} calculates the total intensity for a given emission distribution at the Solar location. For our purpose, we consider a total of 12 solar locations (intervals of $\phi=30^{\circ}$, ($R_\odot$, $\phi$, 0) in cylindrical coordinates)
to account for the unknown azimuthal location of the Sun relative to the star clusters. We choose $\phi=0^{\circ}$ and $\phi=270^{\circ}$ as our default values for DI and DI+CI cases, respectively, but variations with other Solar positions are shown in supplementary material. To calculate the X-ray emissivity at a given temperature, we integrate the \textsc{apec} spectral model (\citealt{Smith2001}) within the R4 band. 
Next, we compare the brightness map for the two runs (with and without central injection). The left panel in Fig. \ref{fig:bubble} shows the SB map at $50$ Myr (so that the eROSITA bubbles' age is $\approx 27$ Myr) using the Aitoff-Hammer projection map for the `DI' run. Clearly, we do not see any structures like the eROSITA bubble; we see smaller structures (than eRBs), which are due to the shocked gas outside the hot chimneys. In the right panel, we show the SB map for the `DI+CI' run, and we see the presence of two large and coherent X-ray structures of size $\sim 80\deg$ at $50$ Myr among the previously seen smaller structures. 
The size, brightness, and sharpness of their outer edge resemble the eROSITA bubbles. 
This clearly demonstrates that \textit{the overlapping hot chimneys from the $\sim 5$ kpc star formation ring are inadequate to create the eRB-like structure in our Galaxy, and we require a strong central injection to have features like eROSITA bubbles.}

We also notice some patchiness in the simulated eRBs. This patchiness arises from the presence of pre-existing shocked and low-density regions that are already present (due to the distributed clusters in the disk) when the central energy injection is initiated. These regions affect the subsequent expansion of the outflow, leading to the asymmetric morphology of the eROSITA bubbles in our simulation. Depending on the 3D location of pre-existing shocks and low-density regions in the Milky Way, it is possible that our Solar position is unique in order to see the bubbles. These bubbles might not be visible or might be only partially visible from other locations. We find that a solar location between $\phi=120^\circ$ and $270^\circ$ shows the eROSITA bubbles (see supplements) but is not clearly visible at other angles. This means about $50$\% of the possible solar locations in the DI+CI case show the eRBs, in comparison to the DI case, which does not show eRB-like features from any vantage angle.

We explored the effect of changing the Solar position along the $\phi$ direction on the resulting X-ray SB maps (fig. \ref{fig:solar_pos_DI1}, \ref{fig:solar_pos_DI2}, \ref{fig:solar_pos_CI1}, \ref{fig:solar_pos_CI2}). No bubble-like morphology was found for any of the Solar positions tested, which provides additional support for a central injection scenario as the origin of the eROSITA bubbles. Since these bubbles are produced by Galactic outflows, they are expected to be enriched in metals. To examine whether enhanced metallicity could produce an eRB-like feature in the X-ray maps, we artificially increased the metallicity of the injected stellar feedback material by factors of 10. However, no bubble-like structures were observed in either case (Fig. \ref{fig:metal_var}), indicating that metallicity enhancement alone cannot account for the formation of the eROSITA bubbles. 

We further investigated the X-ray SB maps across different X-ray bands (i.e., in R1, R2, R5, and R6 bands) and found no evidence of eROSITA bubble-like structures (Fig. \ref{fig:band_varDI} and \ref{fig:band_varCI}). Combined with the previous tests involving the Solar position and metallicity enhancement, these results provide additional support for a central energy injection as the origin of the eROSITA bubbles. 

It would be worthwhile to extend the present simulations by including additional physical processes such as magnetic fields and cosmic rays to conclude about radio emission related to the eRBs and their polarization properties. Such simulations would enable the prediction of associated gamma-ray emission and the construction of synthetic gamma-ray maps, allowing for a direct comparison with observations of the Fermi bubbles as well. A joint analysis of the X-ray and gamma-ray morphology would provide stronger constraints on the physical mechanisms responsible for the formation and evolution of both the eROSITA and Fermi bubbles. We restrict our analyses only to the hydrodynamics due to the available computational resources.

To conclude, in this work, we have investigated the origin of the eROSITA bubbles using three-dimensional hydrodynamical simulations of Galactic outflows, particularly to settle the debate between distributed energy injection and a central energy injection as a possible origin scenario. We have provided ample tests to show that a central energy injection from the Galactic Center is essential to produce the characteristic morphology of the observed eROSITA bubbles. In contrast, outflows driven by a starburst ring at $\sim 5$ kpc in the Milky Way fail to generate similar bubble-like structures, even after exploring variations in the Solar position, metallicity enhancement, and X-ray energy bands. These findings strongly favor a central injection scenario as the primary mechanism responsible for the formation of the eROSITA bubbles.

\begin{acknowledgments}
MSB thanks Karen Yang for the useful discussion.
MSB acknowledges support from the National Science and Technology Council (NSTC) of Taiwan (NSTC 115-2112-M-007-015-MY3).

\end{acknowledgments}

\appendix

\begin{figure*}
    \centering
    \includegraphics[width=0.9\textwidth]{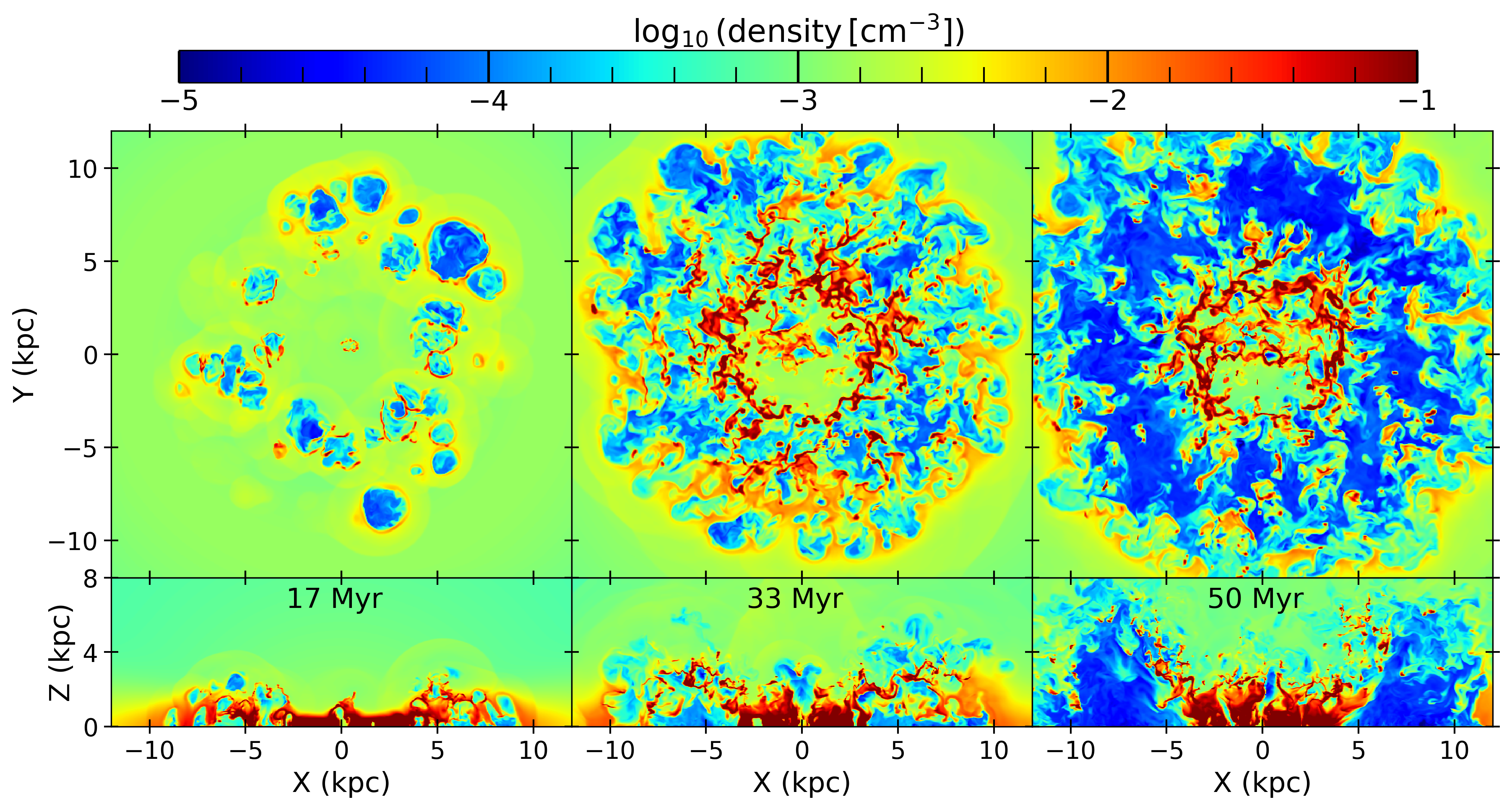}
    \caption{Time evolution of the DI simulation case. Same as Fig. \ref{fig:time_evol_temp_DI}, but for density (DI run).}
    \label{fig:time_evol_density_DI}
\end{figure*}

\begin{figure*}
    \centering
    \includegraphics[width=0.9\textwidth]{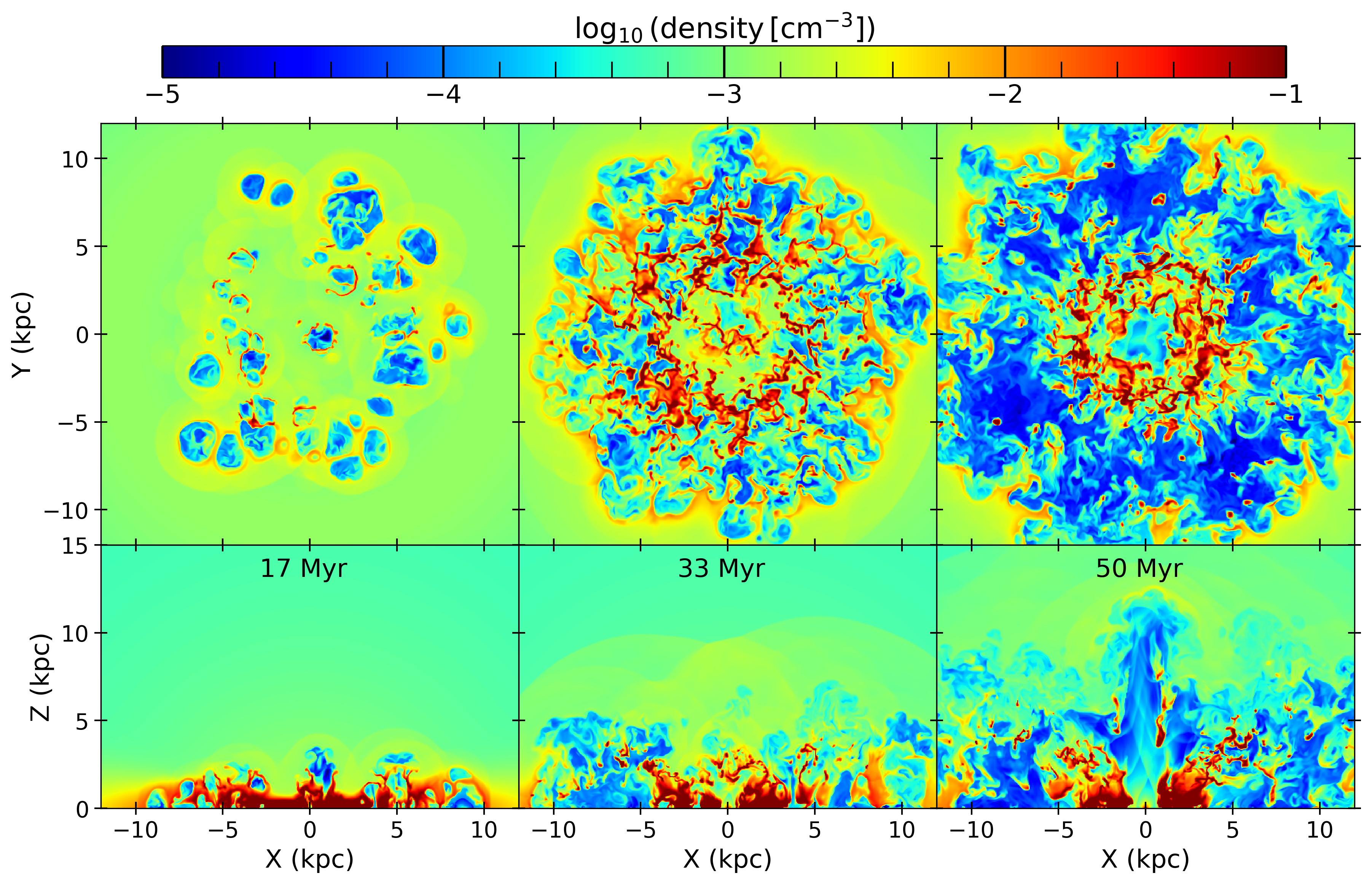}
    \caption{Same as Fig. \ref{fig:time_evol_temp_CI}, but for density (DI+CI run).}
    \label{fig:time_evol_density_CI}
\end{figure*}

\begin{figure*}
    \centering
    \includegraphics[width=0.8\textwidth]{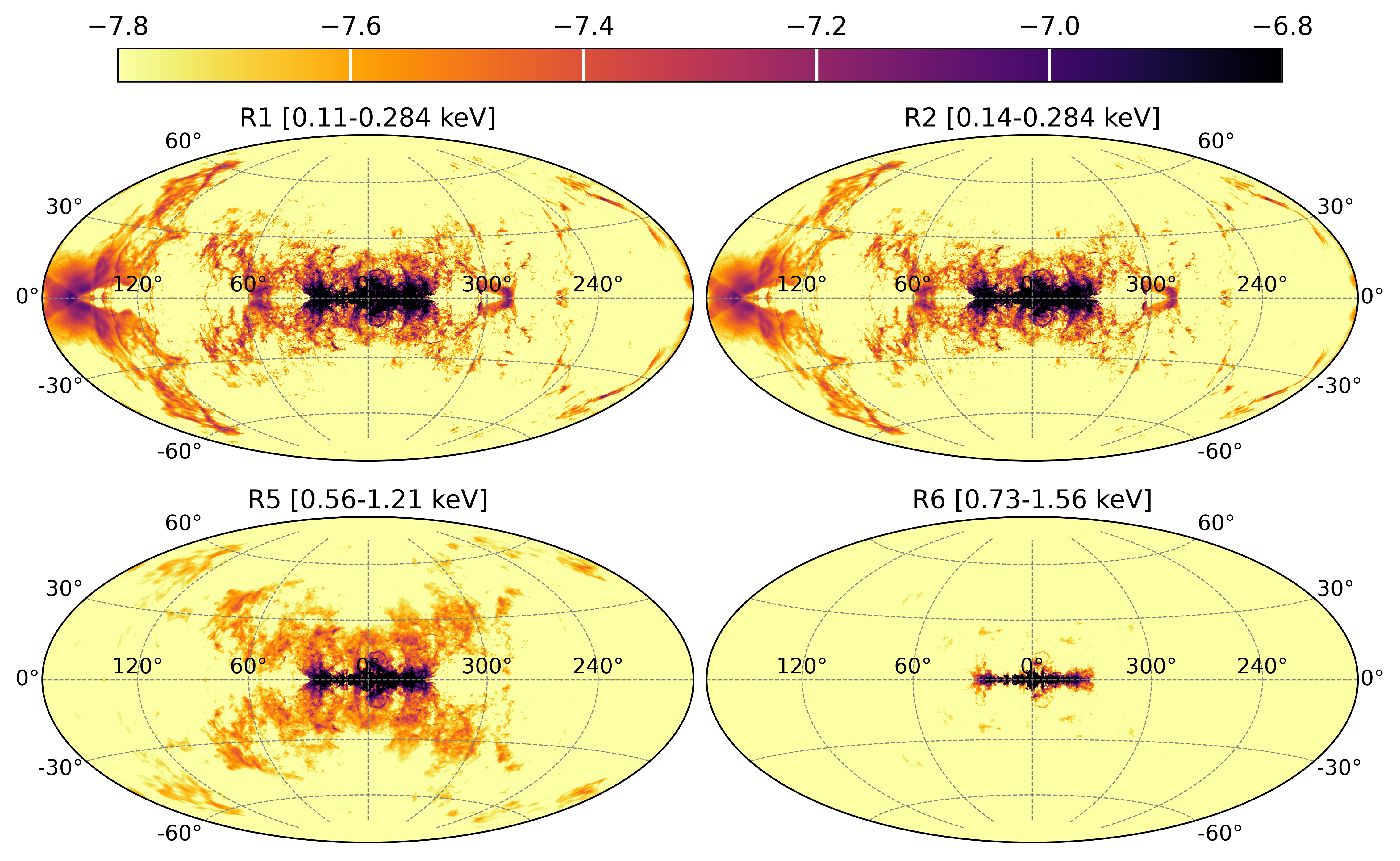}
    \caption{Surface brightness maps in various ROSAT energy bands for the DI case. No feature like the eROSITA bubbles is present in any of these bands. Therefore, our conclusions remain unaffected by the chosen energy bands.}
    \label{fig:band_varDI}
\end{figure*}

\begin{figure*}
    \centering
    \includegraphics[width=0.8\textwidth]{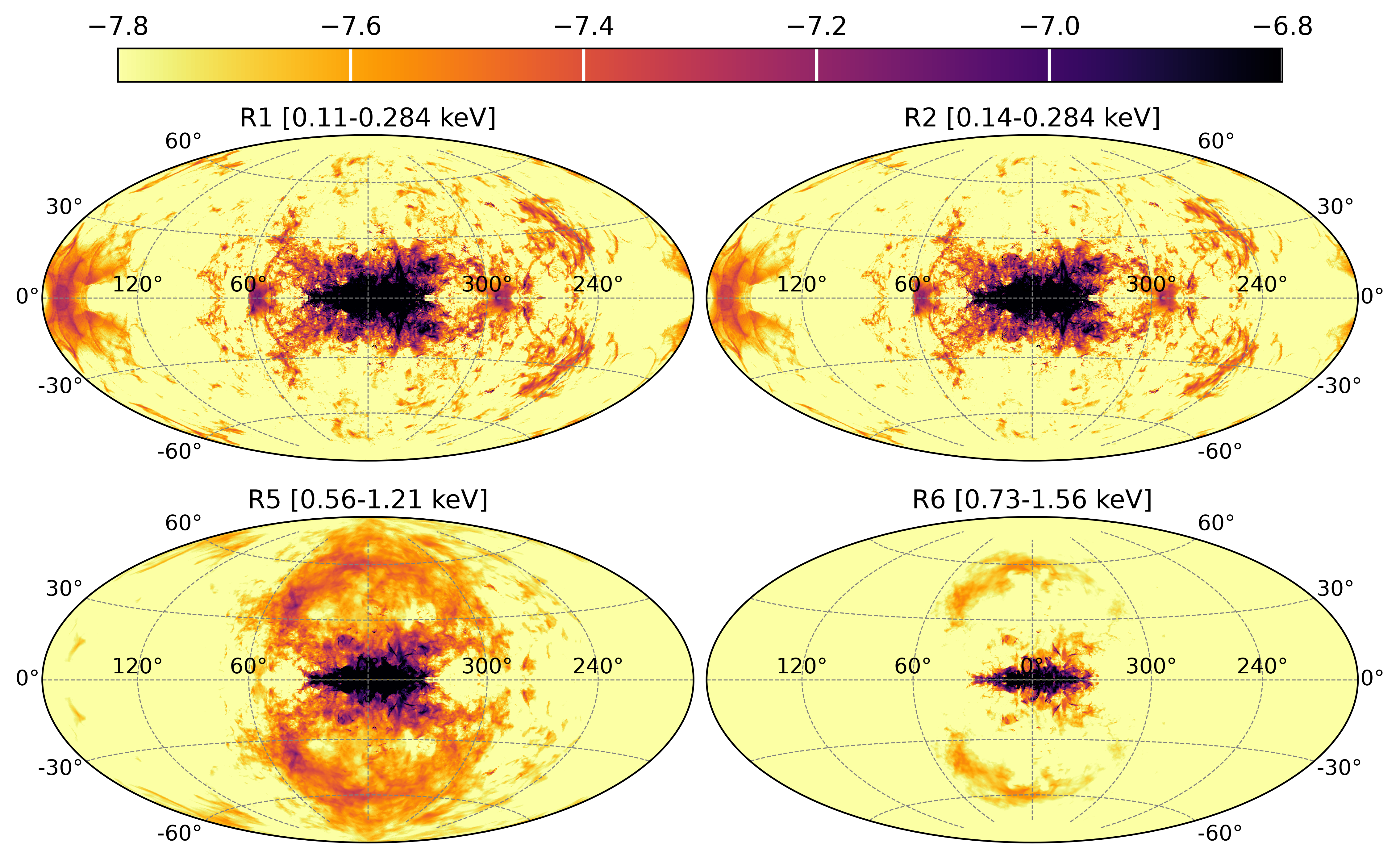}
    \caption{Surface brightness maps in various ROSAT energy bands for the DI+CI case. eROSITA bubbles are not visible in softer bands, but are visible in R5 and R6 bands, along with the R4 band (shown in the next figures).}
    \label{fig:band_varCI}
\end{figure*}

\begin{figure*}
    \centering
    \includegraphics[width=0.9\textwidth]{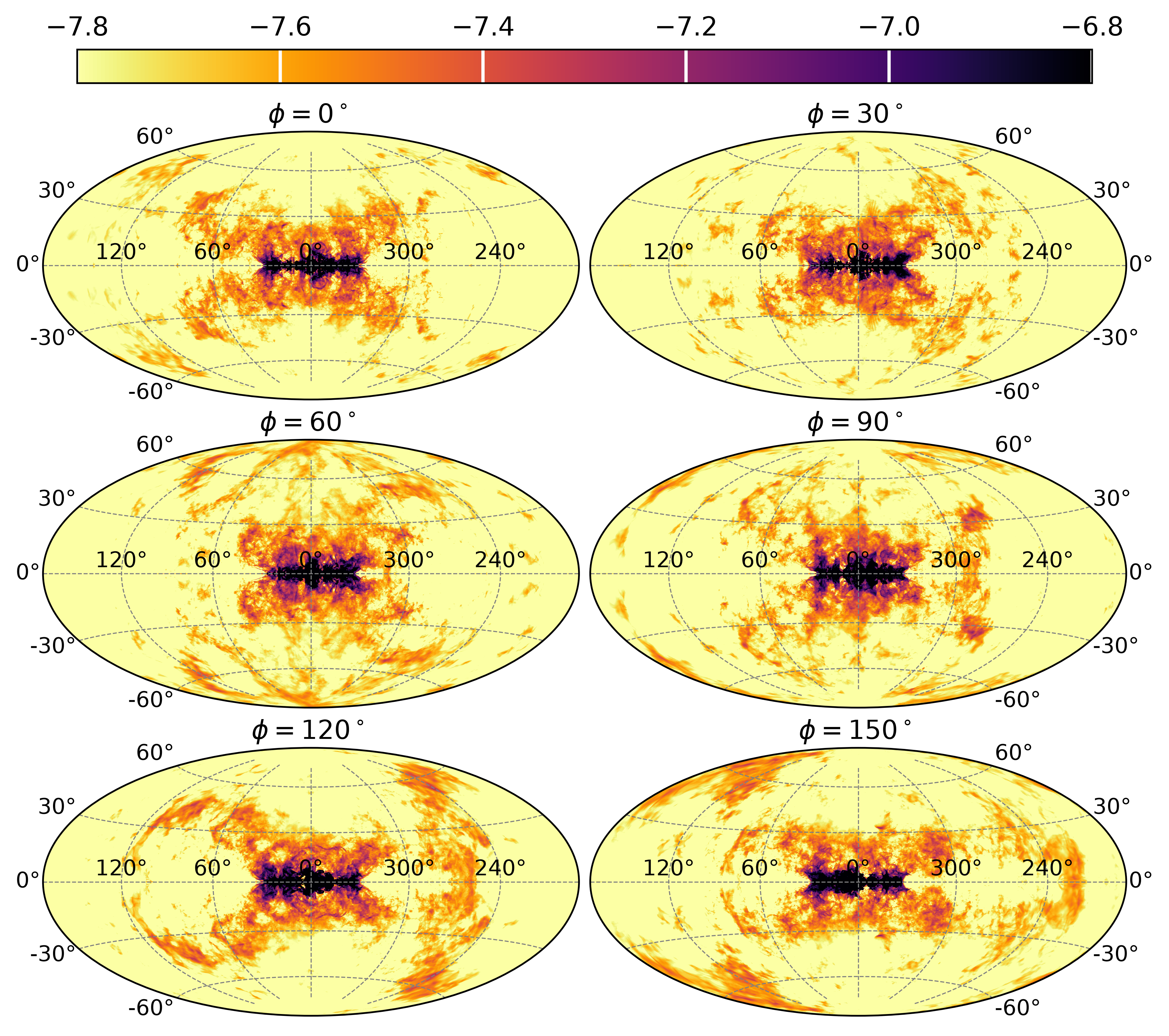}
    \caption{Surface brightness map in R4 band from different Solar positions (($R_\odot$, $\phi$, 0) in cylindrical coordinates) for the distributed injection run. There are no coherent and strong shocked features (resembling the eROSITA bubbles) in any of the hemispheres (east/west). This shows that our results are not affected by the random solar location chosen while creating the SB map.}
    \label{fig:solar_pos_DI1}
\end{figure*}

\begin{figure*}
    \centering
    \includegraphics[width=0.9\textwidth]{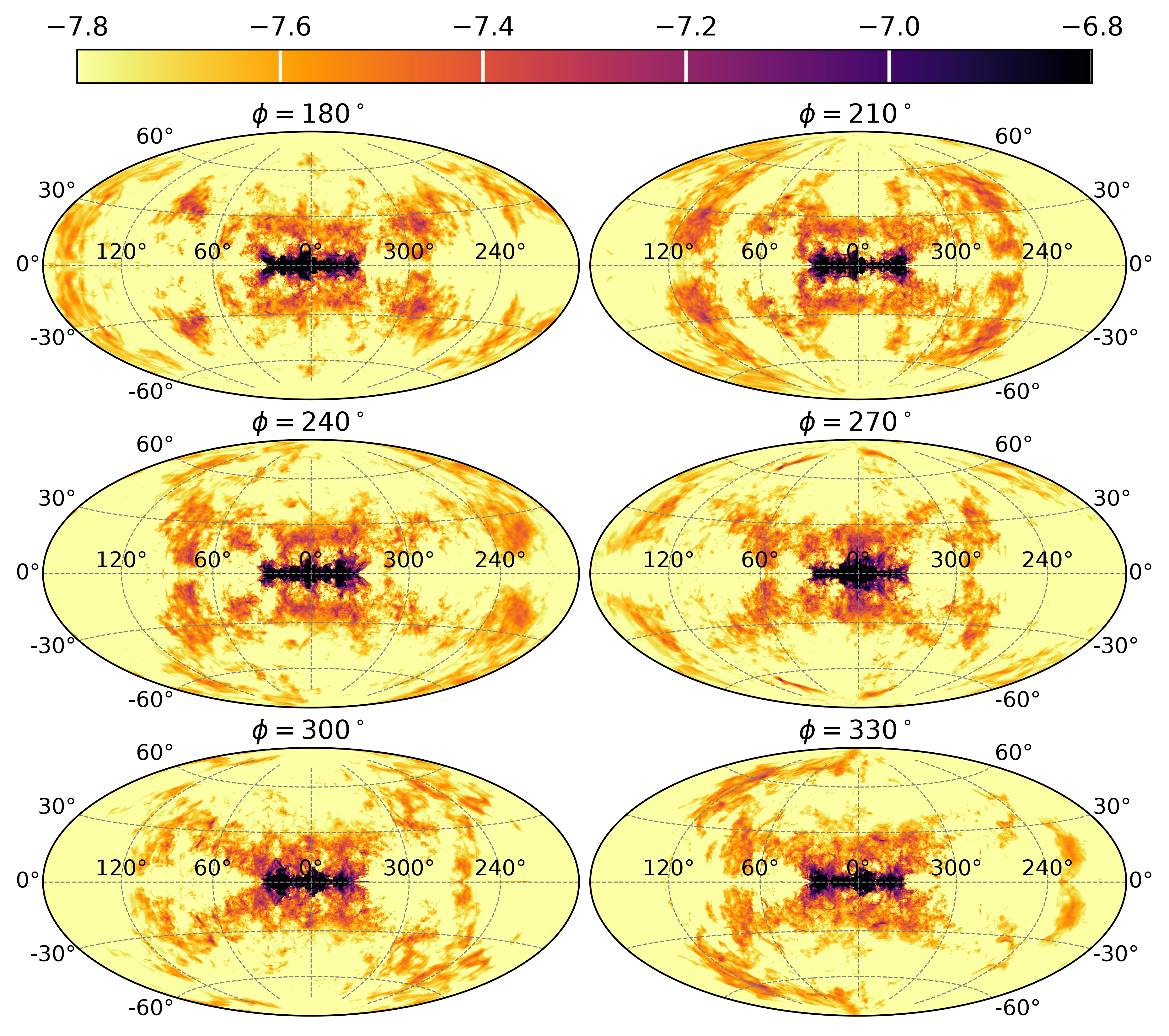}
    \caption{Continuation of Fig. \ref{fig:solar_pos_DI1}. The eROSITA bubble is not seen from any solar position.}
    \label{fig:solar_pos_DI2}
\end{figure*}    

\begin{figure*}
    \centering
    \includegraphics[width=0.9\textwidth]{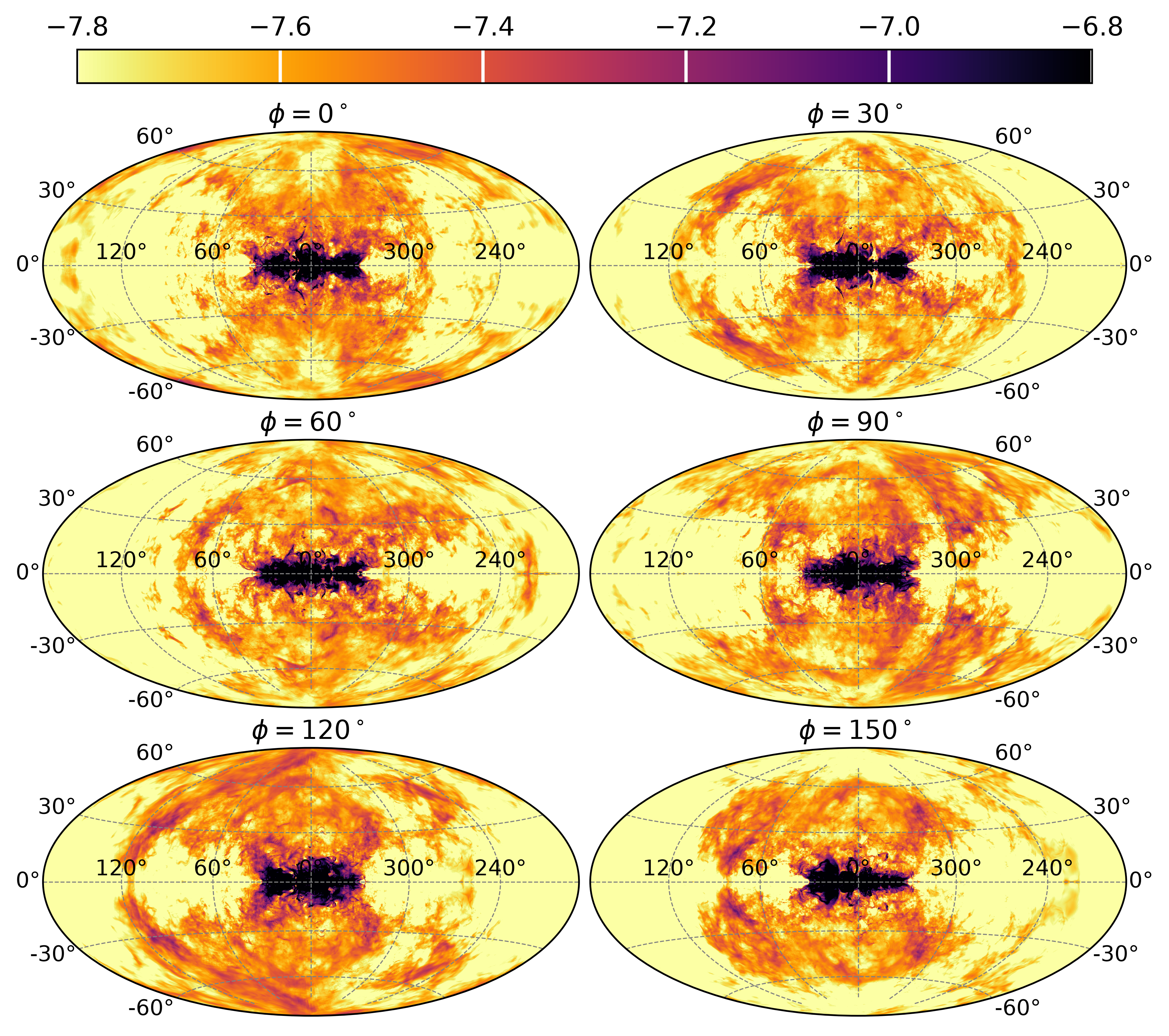}
    \caption{SB map in R4 band for DI+CI case. Different subplots are for different Solar positions, i.e., ($R_\odot$, $\phi$, 0). This set of projections do not show any strong eROSITA-like features, except at $\phi> 120^\circ$.}
    \label{fig:solar_pos_CI1}
\end{figure*}

\begin{figure*}
    \centering
    \includegraphics[width=0.9\textwidth]{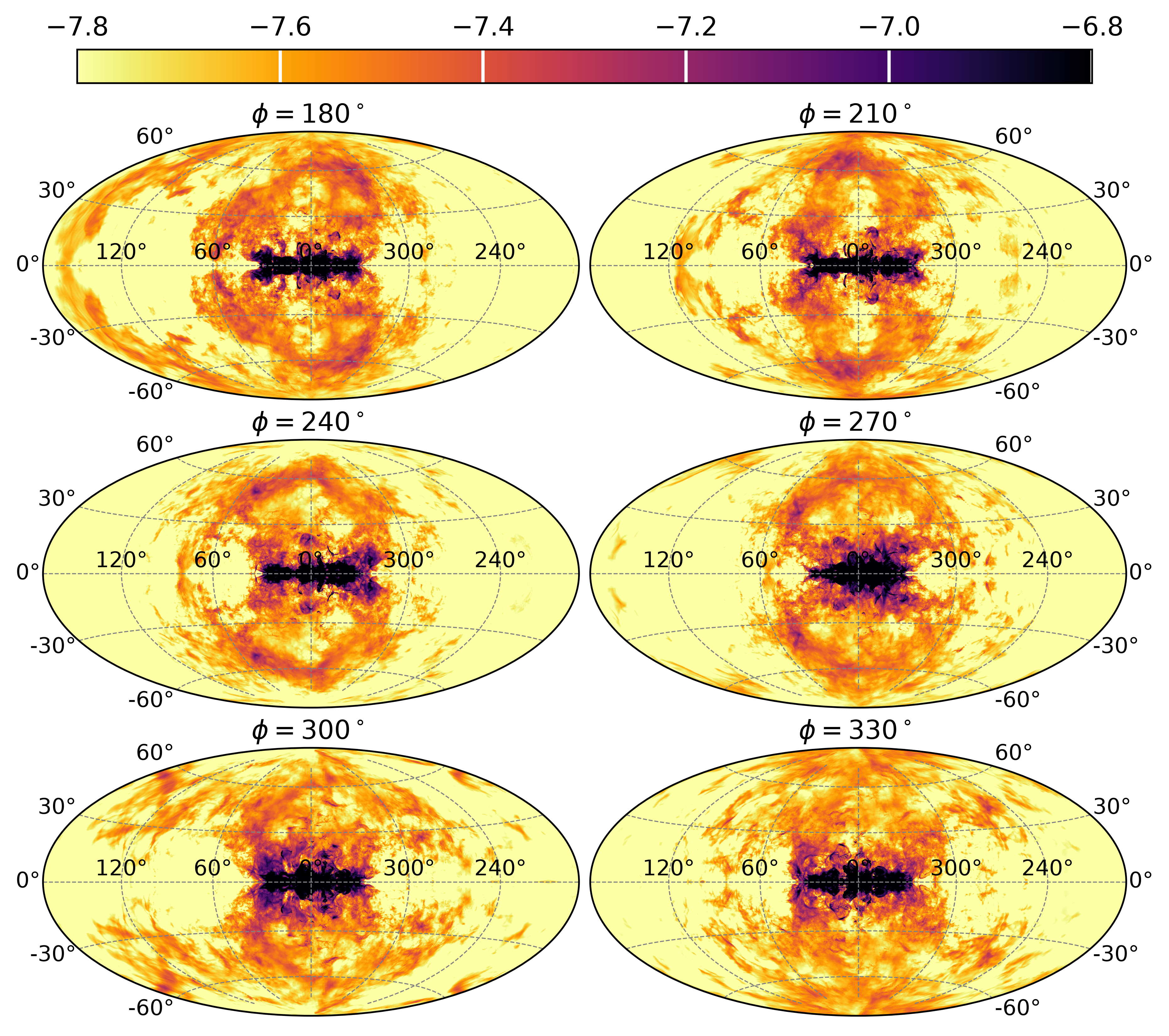}
    \caption{Continuation of Fig. \ref{fig:solar_pos_CI1}. eROSITA bubble is clearly seen for $\phi \gtrsim 120^{\circ}$ and $\phi \lesssim 270^{\circ}$, but the bubble looks more prominent at $\phi \approx 270^\circ$.}
    \label{fig:solar_pos_CI2}
\end{figure*}


\bibliography{review}{}
\bibliographystyle{aasjournalv7}



\end{document}